\UseRawInputEncoding

\documentclass{cpbtex}
\usepackage{dsfont}
\usepackage{graphicx}
\usepackage{dcolumn}
\usepackage{bm}
\usepackage{setspace}
\usepackage{color}
\usepackage{lineno}
\usepackage[T1]{fontenc}
\begin{document}
\begin{CJK*}{GBK}{song}

\title{Facilitation of controllable excitation in Rydberg atomic ensembles\thanks{The National Science Foundation of China (Grants No. 12174106, No. 11474094), and the Science and Technology Commission of Shanghai Municipality (Grant No. 18ZR1412800).}}

\author{Han Wang, Jing Qian\thanks{E-mail:~jqian1982@gmail.com}\\
{State Key Laboratory of Precision Spectroscopy, Department of Physics, School of Physics}\\
{and Electronic Science, East China Normal University, Shanghai 200062, China}} 
\date{\today}
\maketitle
\begin{abstract}
    Strongly-interacting Rydberg atomic ensembles have shown intense collective excitation effects due to the inclusion of single Rydberg excitation shared by multiple atoms in the ensemble. In this paper we investigate a counter-intuitive Rydberg excitation facilitation with a strongly-interacting atomic ensemble in the strong probe-field regime, which is enabled by the role of a control atom nearby. Differing from the case of a single ensemble, we show that, the control atom's excitation adds to a second two-photon transition onto the doubly-excited Rydberg state, arising an excitation facilitation for the ensemble atoms. Our numerical studies depending on the method of quantum Monte Carlo wavefunction, exhibit the observation constraints of this excitation facilitation effect under practical experimental conditions. The results obtained can provide a flexible control for the excitation of Rydberg atomic ensembles and participate further uses in developing mesoscopic Rydberg gates for multiqubit quantum computation.
\end{abstract}

\textbf{Keywords:} Excitation facilitation; Rydberg atom; Monte Carlo Wave Function;  many-body system 

\textbf{PACS:} 33.80.Rv, 32.80.Qk, 05.10.Ln, 73.20.Mf, 45.50.Jf

\section{Introduction}
Collective excitation enhancement associated with strongly-interacting Rydberg atoms constitutes the basis for versatile applications in observing {\it e.g.} many-body effects \cite{PhysRevLett.114.203002,PhysRevA.96.041602,PhysRevLett.123.203603,PhysRevResearch.2.043339}, large-scale quantum computation \cite{PhysRevLett.99.260501,Zhang_2021,Wu_2021,Chinese Phys.B.29.083202}, quantum entanglement \cite{PhysRevA.98.043836,Yang2022,PhysRevLett.128.060502}, generation of single photons \cite{Yang:22,PhysRevResearch.3.033287}, collective emission of photons \cite{PhysRevA.103.023703}, {\it etc}.
The manipulation of a strongly-interacting atomic ensemble can enable the implementation of single ensemble qubit gate \cite{PhysRevA.102.042607}, a many-particle GHZ state \cite{PhysRevA.98.062326, PhysRevResearch.4.033087,RevModPhys.82.2313}, a photon reflection phase \cite{Motzoi_2018}, and the creation of control-ensemble entanglement \cite{PhysRevLett.102.170502}.
So far, many of these achievements depend on the Rydberg blockade mechanism since it causes a single atom's excitation shared by $N$ ensemble atoms within the range of blockade radius \cite{PhysRevLett.99.163601,PhysRevA.87.053414,PhysRevX.5.031015,PhysRevLett.128.123601}. This fully-blockaded atomic ensemble (or so-called a {\it superatom}) can be described by a reduced three-level structure using symmetric Dicke states \cite{PhysRevLett.107.213601}, and benefits from a $\sqrt{N}$-enhancement of the weak probe strength in the collective excitation environment \cite{Dudin2012,PhysRevA.87.023401,PhysRevA.97.043811}.

Although the {\it superatom} model could fundamentally exhibit the mechanism of Rydberg blockade, it only works with a perfect eletromagnetically induced transparency condition where the probe strength is much weaker than that of the coupling one \cite{PhysRevLett.105.193603}. A stronger probe field will lead to unexpected multi-photon processes associated with various intermediate and Rydberg states, which renders the incoherent dissipative decays onto other asymmetric collective states non-negligible \cite{Qiao_2021,PhysRevA.105.043715}. It is therefore important to completely capture the feature of collective Rydberg excitation with an {\it improved superatom model}, which is beyond the weak probe regime \cite{PhysRevA.89.033839}.

Motivated by a pioneering work in \cite{PhysRevLett.113.233002}, G\"{a}rttner and co-authors discovered a novel collective excitation enhancement effect in the weak probe regime because of the multi-photon transition among collective Dicke states. In the present work, we extend this mechanism by investigating the collective excitation enhancement of Rydberg-ensemble atoms with the help of a control atom nearby. This enhancement effect is found to occur beyond the weak probe regime where the probe strength is comparable to or larger than the coupling strength \cite{Bai_2018}.
We show that, due to the use of a co-excited control atom, the coupled control-ensemble system would obtain an auxiliary excitation channel towards the doubly-excited Rydberg state. When this state is adjusted to be resonantly coupled via an antiblockade facilitation, the steady-state Rydberg population will redistribute between the singly- and doubly-excited Rydberg states, exhibiting an apparent excitation enhancement effect \cite{PhysRevLett.98.023002,PhysRevLett.104.013001,Kara2018,Bai_2020}. By utilizing the quantum Monte Carlo wavefunction method \cite{Petrosyan_2013}, we numerically verify that, this facilitated excitation can persistently exist with strong probe drivings. By using realistic parameters from experimental setups, we show the
regimes where the excitation facilitation effect can happen as long as the temperature of atoms and the control-ensemble distance are appropriately determined. 
\section{\label{sec:level1}Theoretical Strategy}
\subsection{Atomic ensemble and Dicke states}
\begin{figure}
\centering\includegraphics[scale=0.55]{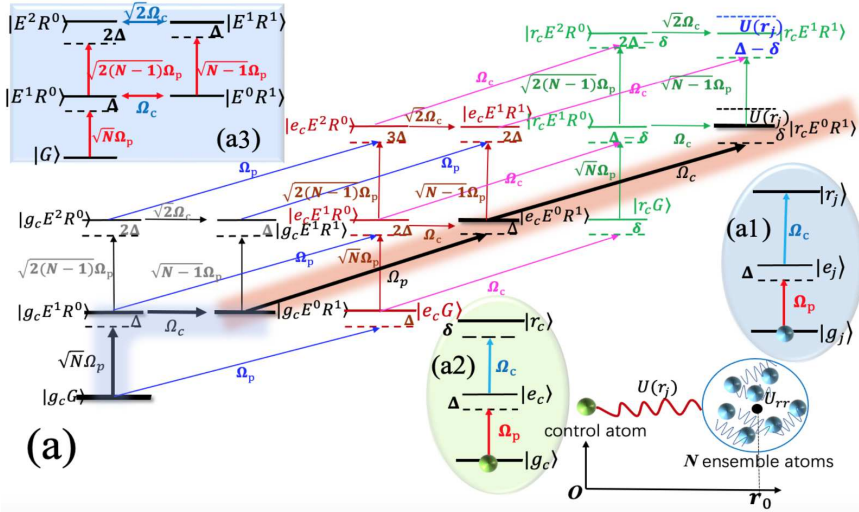}
\caption{Controllable excitation facilitation in a coupled control-ensemble system. (a) Full level scheme of a control atom $\{|g_c \rangle,|e_c \rangle,|r_c \rangle\}$ [see (a2)] interacting with a strongly-blockaded Rydberg atomic ensemble which is described by five symmetric collective states $\{|G\rangle,|E^1 R^0 \rangle,|E^0 R^1 \rangle,|E^2 R^0 \rangle,|E^1 R^1 \rangle\}$ [see (a3)]. The cascade excitation obeying the transition of $|g_cG \rangle\to|g_cE^1R^0 \rangle\to|g_cE^0R^1 \rangle\to|e_cE^0R^1 \rangle\to|r_cE^0R^1 \rangle$(marked by the blue-shaded and red-shaded areas), is facilitated by an exact antiblockade condition $\delta = -U(\bm{r}_j)$ with respect to the doubly-excited Rydberg state $|r_cE^0R^1\rangle$. Insets: Level schemes and atom-light couplings with (a1) the $j$th ensemble atom, (a2) the control atom and (a3) the atomic ensemble.}
\label{fig1:model}
\end{figure}

We consider a strongly-interacting atomic ensemble within the blockade volume which is composed by $N$ three-level atoms [see Fig.\ref{fig1:model}(a1)]. For the $j$th ensemble atom, the ground state $|g_j\rangle$ is off-resonantly coupled to an intermediate state $|e_j\rangle$ with Rabi frequency $\Omega_p$ and frequency detuning $\Delta$. A second laser resonantly drives the subsequent transition between $|e_j\rangle$ and a Rydberg level $|r_j\rangle$ with Rabi frequency $\Omega_c$. The strong intraspecies interactions $U_{rr}$ would lead to a perfect dipole blockade effect that accommodates just one Rydberg excitation here \cite{PhysRevLett.87.037901,doi:10.1126/science.1217901}. Also, the Hamiltonian is invariant under the exchange of particles since all the ensemble atoms are identical. Therefore the atomic ensemble can be described by a set of collective symmetric Dicke states $\{|G\rangle,|E^mR^0\rangle,|E^{m-1}R^1\rangle\}$ \cite{Carmele_2014}, in which
\begin{equation}
|G\rangle=|g_1,g_2,...,g_N\rangle
\end{equation}
stands for all atoms being in the ground state. When the probe field is not so strong it is reasonable to assume at most one ensemble atom can be pumped into intermediate state $|e_j\rangle$ via a one-photon excitation, arising
\begin{equation}
   |E^{1}R^{0}\rangle=\frac{1}{\sqrt{N}}(\sum_{j=1}^{N}|e_j\rangle  \langle g_j|) |G\rangle.
\end{equation}
However for a stronger probe field with more photons inside the ensemble, other states like $|E^2R^0\rangle,|E^3R^0\rangle,|E^4R^0\rangle ...$ are also essential \cite{PhysRevA.87.023827}. We have to consider a general form as 
\begin{equation}
   |E^{m}R^{0}\rangle=\frac{1}{\sqrt{N_e^m}}(\sum_{j=1}^{N}|e_j\rangle  \langle g_j|)^m |G\rangle
\end{equation}
with $N_e^m=N!m!/{(N-m)!}$. State $|E^{m}R^{0}\rangle$ indicates $m$ atoms in the bare state $|e_j\rangle$ and $m\in[0,N]$. Additionally, state
\begin{equation}
    |E^{m-1}R^{1}\rangle=\frac{1}{\sqrt{N_r^m}}(\sum_{j=1}^{N}|e_j\rangle  \langle g_j|)^{m-1}(\sum_{j=1}^{N}|r_j\rangle  \langle g_j|) |G\rangle 
\end{equation}
with $N_r^m=N N!(m-1)!/(N-m+1)!$, means one atom being excited into the Rydberg state $|r_j\rangle$ while others are either in the ground or intermediate states. 
If $m=0$, $|E^{m-1}R^{1}\rangle$ is not present.
Here we treat the targeted atomic ensemble represented by symmetric Dicke states that accommodates only one Rydberg excitation due to the Rydberg blockade, so state $|E^{m-2}R^2\rangle$ is not present.

In addition, note that under the condition of $\Delta\gg\Omega_p$ (the probe field is not so strong), 
the transitions into
states $|E^{> 1}R^1\rangle$ and $|E^{> 2}R^0\rangle$ are also negligible \cite{PhysRevA.86.023828}. See Fig.\ref{fig2:steady}a(inset),
we have
numerically verified that the probability of multi-excitation onto these states is always below $10^{-5}$. Therefore, the atomic ensemble can be safely described by five collective states $\{|G\rangle ,|E^{1}R^{0}\rangle, |E^{0}R^{1}\rangle, |E^{2}R^{0}\rangle, |E^{1}R^{1}\rangle\}$ [Fig.\ref{fig1:model}(a3)], where the corresponding transition coefficients
as shown in the figure, depend on the number of 
atoms $N$ in the ensemble. Based on such an atomic ensemble, a collective excitation enhancement effect has been observed due to the coherent multi-photon couplings among collective states \cite{PhysRevLett.113.233002}. While this finding occurs only in the weak probe regime $0.15\leq\Omega_p/\Omega_c\leq0.4$ where a small $\Delta$ value is required.

\subsection{Control-ensemble system}

Now we introduce another three-level  control atom $\{|g_c
\rangle,|e_c\rangle,|r_c\rangle \}$, individually addressed near the atomic ensemble [Fig.\ref{fig1:model}(a2)]. 
In this manner, single control atom and a mesoscopic atomic ensemble can be confined in two separate 
optical tweezers \cite{doi:10.1080/23746149.2022.2064231}. The way to load an optical tweezer with a desired number of atoms can adopt the probabilistic loading technique \cite{PhysRevLett.96.063001}. Here we assume that the ensemble atoms(blue) distribute with a random position $\bm{r}_j$ while the auxiliary control atom(green) is placed at the zero point $O$ with a distance $\bm{r}_0$ apart from the ensemble center. The control atom features a same three-level structure, and is simultaneously driven by $\Omega_p$ and $\Omega_c$ on the transitions  $|g_c\rangle\to|e_c\rangle$ and $|e_c\rangle\to|r_c\rangle$. Let us now discuss the concrete realization of the excitation facilitation mechanism.

Taking account of all atom-light couplings and interatomic interactions, the full level scheme as illustrated in Fig.\ref{fig1:model}(a), has totally 15 collective states. By assuming a sufficiently large detuning $\Delta\gg\Omega_p,\Omega_c$, numerical estimation for the probability onto these off-resonance energy levels(marked by thin lines) is only $7.13\times10^{-4}$ when $N=3$, $\Omega_p=2\Omega_c$, which can be safely eliminated. So we pay attention to merely five collective states(marked by thick lines),
\begin{equation}
    \{|g_cG\rangle,|g_cE^1R^0\rangle,|g_cE^0R^1\rangle,|e_cE^0R^1\rangle,|r_cE^0R^1\rangle\}.
    \label{trr}
\end{equation}
From Fig.\ref{fig1:model}(a), we know that if the control atom is unexcited, {\it i.e.} it is at $|g_c\rangle$, only states $|g_cG\rangle$, $|g_cE^1R^0\rangle$, $|g_cE^0R^1\rangle$ (blue shaded area) can compose a resonant two-photon transition mediated by a large detuning $\Delta$ to the intermediate $|g_cE^1R^0\rangle$ state. In this case the lower transition between $|g_cG\rangle$ and $|g_cE^1R^0\rangle$ is enhanced by $\sqrt{N}\Omega_p$ \cite{PhysRevLett.127.063604}. As a result, the steady-state Rydberg population $f_r$($=P_{g_c}(\infty)$, see definition in Eq.(\ref{ex})) of ensemble atoms, equivalent to the Rydberg population on state $|g_cE^0R^1\rangle$, can be simply described by 
$f_r=N\Omega_p^2/(N\Omega_p^2+\Omega_c^2)$, 
as same as the prediction by the {\it superatom} model \cite{PhysRevLett.113.233002}.
While remarkably, accounting for the use of a control-ensemble system we find that the excitation of a simultaneously-driven control atom, adds to a subsequent two-photon transition which obeys
\begin{equation}
|g_cE^0R^1\rangle\to|e_cE^0R^1\rangle\to|r_cE^0R^1\rangle.
\end{equation}
 When the detuning $\delta$ with respect to $|r_c E^0 R^1\rangle$ can be overcome by an appropriate control-ensemble interaction $U(\bm{r}_j)$, the steady-state Rydberg population of ensemble atoms described by the total population on states $|g_c E^0 R^1\rangle$ and $|r_c E^0 R^1\rangle$ would be enhanced, which is irrelated to the status of the control atom. Meanwhile the population on the intermediate state $|e_c E^0 R^1\rangle$ will be suppressed due to a large detuning $\Delta$. 

 \subsection{Imperfect antiblockade condition}

Ideally we expect an exact two-photon resonance $\delta=-U(\bm{r}_j)$(antiblockade) for each ensemble atom, which makes the double Rydberg state $|r_cE^0R^1\rangle$ resonantly-coupled \cite{Ding_2019,PhysRevA.82.062328,PhysRevA.101.042328}.
However accounting for the thermal motion of atoms the real control-ensemble interaction $U(\bm{r}_j)$ must be space-dependent. The thermal distribution of ensemble atoms should be characterized by a Gaussian function \cite{PhysRevLett.118.063606}
\begin{equation}
    f(\bm{r}_j) = e^{-\frac{(\bm{r}_j-\bm{r}_0)^2}{2\sigma^2}}
    \label{Gau1}
\end{equation}
with its width $\sigma=\sqrt{k_BT/m\omega^2}$ where $T$, $m$, $\omega$ are the atomic temperature, atomic mass and trap frequency respectively. Correspondingly, we obtain a fluctuated dipolar interaction \cite{PhysRevLett.128.013603}
\begin{equation}
    U(\bm{r}_j) \approx U(\bm{r}_0)+\delta U(\bm{r}_j).
    \label{perU}
\end{equation}
where $U(\bm{r}_0)=\frac{C_6}{|\bm{r}_0|^6}$ with $C_6$ the interaction coefficient for state $|r_cr_j\rangle$ and $\bm{r}_0$ the displacement between the ensemble center and the control atom. 
The small fluctuated interaction shift is
\begin{equation}
    \delta U(\bm{r}_j)=-\frac{6 C_6(|\bm{r}_j-\bm{r}_0|)}{|\bm{r}_0|^7}
    \label{devr}
\end{equation}
caused by a random displacement $\bm{r}_j$ of atoms obtained from the function $f(\bm{r}_j)$. As a consequence we set $\delta=-U(\mathbf{r}_0)$ in the calculation. For a sufficiently low temperature, the exact antiblockade condition $\delta\approx-U(\bm{r}_j)$ is approximately satisfied since the atoms are frozen and $\delta U(\bm{r}_j)\to 0$. Whereas
the fluctuated component $\delta U(\bm{r}_j)$($\propto  |\bm{r}_j-\bm{r}_0|$) becomes larger if the temperature grows leading to the breakdown of antiblockade facilitation (see more details in Sec.5.2).
Our numerical results depend on a fluctuated interaction $U(\bm{r}_j)$.

\section{Hamiltonian and Numerical method\label{Hami}}

The associated Hamiltonian of the control-ensemble scheme can be written as
\begin{equation}
    \mathcal{H} = \mathcal{H}_{en} + \mathcal{H}_c +  \sum_{j=1}^N U(\bm{r}_j)|r_c\rangle\langle r_c|\otimes|r_j\rangle \langle r_j|
    \label{total}
\end{equation}
where the first term
\begin{equation}
  {\mathcal{H}}_{en}=\sum_{j=1}^{N}\mathcal{H}_0^{(j)}+\sum_{k>j}^NU_{rr}(\bm{r}_j,\bm{r}_k)|r_j\rangle\langle r_j|\otimes|r_k\rangle\langle r_k|
  \label{ensemble}
\end{equation}
represents the Hamiltonian of an atomic ensemble. For each ensemble atom $j$, the single-atom Hamiltonian $\mathcal{H}_0^{(j)}$ reads
\begin{equation}
{\mathcal{H}}_{0}^{(j)}=-\Delta|e_j\rangle\langle e_j|+(\Omega_p|e_j\rangle\langle g_j|+\Omega_c|r_j\rangle\langle e_j|+\text{H.c.}).  
\end{equation}

Besides, we assume the intraspecies interaction $U_{rr}(\bm{r}_j,\bm{r}_k)$
is also strong that leads to a perfect blockade effect within the ensemble atoms, where $\bm{r}_j$ and $\bm{r}_k$ are the random atomic positions. The second term $\mathcal{H}_c$ describing the Hamiltonian of single control atom, can be described by
\begin{equation}
\mathcal{H}_{c}=-\Delta\vert e_c\rangle\langle e_c\vert+\delta\vert r_c\rangle\langle r_c\vert+(\Omega_p\vert e_c\rangle\langle g_c\vert + \Omega_c \vert r_c\rangle\langle e_c\vert +\text{H.c.})
\label{coneq}
\end{equation}
and the third term means the control-ensemble interaction, see Eq.(\ref{perU}).

To our knowledge, exact treatment with a master equation method without mean-field approximation is rather computationally-demanding \cite{PhysRevA.98.062714}, especially for a many-body system containing complex interactions. Below we 
numerically solve the time-dependent evolution with the Monte Carlo wave-function method \cite{PhysRevA.91.043402}. In the simulation, state $|\Psi\rangle$ of system evolves according to a stochastic Schr$\ddot{o}$dinger equation
\begin{equation}
    \partial_t\vert\Psi\rangle=-i(\mathcal{H} - \frac{i}{2}\mathcal{L}^2)\vert \Psi\rangle
\label{stochastic}
\end{equation}
where the Liouvillian dissipative operator serving as a non-Hermitian term, reads
\begin{equation}
 \mathcal{L}^2=\sum_{j=1}^N(\hat{{\mathcal{L}}}_e^{j\dagger}\hat{{\mathcal{L}}_e^j}+\hat{{\mathcal{L}}}_r^{j\dagger}\hat{{\mathcal{L}}_r^j}+\hat{{\mathcal{L}}}_{z1}^{j\dagger}\hat{{\mathcal{L}}_{z1}^j}+\hat{{\mathcal{L}}}_{z2}^{j\dagger}\hat{{\mathcal{L}}}_{z2}^j) 
 +(\hat{{\mathcal{L}}}_e^{c\dagger}\hat{{\mathcal{L}}_e^c}+\hat{{\mathcal{L}}}_r^{c\dagger}\hat{{\mathcal{L}}_r^c}+\hat{{\mathcal{L}}}_{z1}^{c\dagger}\hat{{\mathcal{L}}_{z1}^c}+\hat{{\mathcal{L}}}_{z2}^{c\dagger}\hat{{\mathcal{L}}}_{z2}^c). \label{diss}
\end{equation}
Following Eq.(\ref{diss}), the whole dissipation process in the control-ensemble system contains three parts: 

(i) The spontaneous decay from the intermediate state $\vert e_{j(c)}\rangle$ with rate $\Gamma_e$ is denoted by
\begin{equation}
\hat{{\mathcal{L}}}_{e}^{j(c)}=\sqrt{\Gamma_{e}}|g_{j(c)}\rangle\langle e_{j(c)}|, 
\end{equation}

(ii) the spontaneous decay from the Rydberg state $\vert r_{j(c)}\rangle$ with rate $\Gamma_r$ reads
\begin{equation}
\hat{{\mathcal{L}}}_{r}^{j(c)}=\sqrt{\Gamma_{r}}|g_{j(c)}\rangle\langle r_{j(c)}|.   
\end{equation}

(iii) the dephasing effect with rates $\gamma_{ge}$ and $\gamma_{er}$ caused by atomic motions and collisions takes form of
\begin{eqnarray}
\hat{{\mathcal{L}}}_{z1}^{j(c)}=\sqrt{\gamma_{ge}}(|e_{j(c)}\rangle\langle e_{j(c)}|-|g_{j(c}\rangle\langle g_{j(c)}|)   \nonumber\\
\hat{{\mathcal{L}}}_{z2}^{j(c)}=\sqrt{\gamma_{er}}(|r_{j(c)}\rangle\langle r_{j(c)}|-|e_{j(c)}\rangle\langle e_{j(c)}|).
\end{eqnarray}
In each quantum trajectory $m$, the evolution of $|\Psi_m\rangle$ obeys the stochastic Schr$\ddot{o}$dinger equation (\ref{stochastic}) \cite{PhysRevA.85.063822}. However it interrupts via a random quantum jump $\vert \Psi_m\rangle \to \hat{\mathcal{L}}_{a}^{k}\vert \Psi_m\rangle$ with $a\in\{e,r,z1,z2\}$ and $k\in\{j,c\}$, determined by its weight $dt \langle\Psi_m\vert\hat{\mathcal{L}}_{a}^{k\dagger}\hat{\mathcal{L}}_{a}^{k}\vert\Psi_m\rangle$. At the same time $|\Psi_m\rangle$ should be re-normalized via $\vert\bar{\Psi}_m(t)\rangle=\vert\Psi_m(t)\rangle/\sqrt{\langle\Psi_m(t)\vert\Psi_m(t)\rangle}$ for keeping the preservation of population \cite{PhysRevLett.108.023602}. Finally the calculated density matrix $\bar{\rho}(t)$ is obtained by averaging over sufficient quantum trajectories $M=300$(used), which takes form of
\begin{equation}
  \bar{\mathcal{\rho}}(t)=\frac{1}{M}\sum^{M}_{m=1}|\bar{\Psi}_m(t)\rangle\langle\bar{\Psi}_m(t)|.
\end{equation}

The population dynamics of arbitrary Rydberg-excited states in the control-ensemble system can be written as 
\begin{equation}
  P_{s}(t)=\text{Tr}[\bar{\rho}\hat{\mathbf{O}}_{s}]=\frac{1}{M}\sum_{m=1}^M\langle\bar{\Psi}_m(t)|\hat{\mathbf{O}}_{s}|\bar{\Psi}_m(t)\rangle 
  \label{ex}
\end{equation}
where the operator is
\begin{eqnarray}
  \hat{\mathbf{O}}_{s} &=& |s\rangle\langle s|\otimes\{\sum_{j=1}^{N}|r_j\rangle\langle r_j|\prod_{k\neq j}^{N}(|g_k\rangle\langle g_k|)\}
\end{eqnarray}
with $s\in(g_c,r_c)$ denoting the state of the control atom. $P_{g_c}(t)$ or $P_{r_c}(t)$ stand for the time-dependent Rydberg population of the targeted ensemble when the control atom is unexcited or excited. Note that we treat $P_{g_c}(t)+P_{r_c}(t)$ as the Rydberg excitation probability of ensemble atoms. Replacing the operator $\hat{\mathbf{O}}_{s}$ by
\begin{equation}
 \hat{\mathbf{O}}_{g_cG} = |g_c\rangle\langle g_c|\otimes\prod_{j=1}^{N}(|g_j\rangle\langle g_j|)  
\end{equation}
in Eq.(\ref{ex}) arises a new expression $P_{g_cG}(t)$ which stands for the population dynamics of the ground state $|g_cG\rangle$.

\begin{figure}
\centering
\includegraphics[scale=0.40]{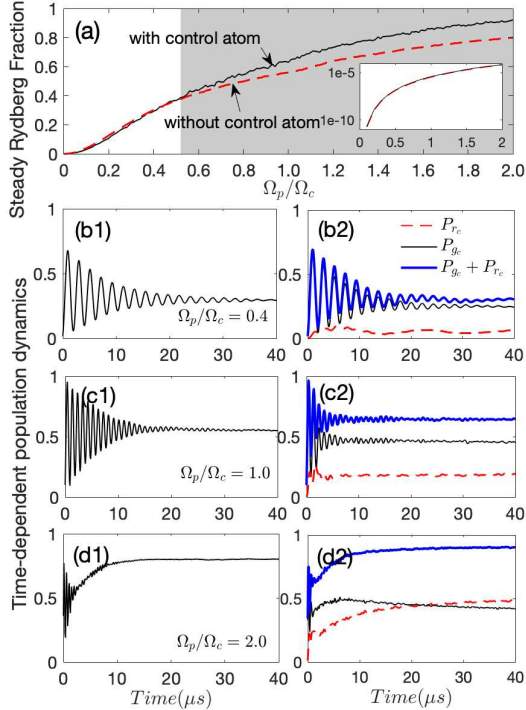}
\caption{(a) Steady-state Rydberg population $f_r$ of the targeted atomic ensemble {\it vs.} the variation of $\Omega_p/\Omega_c$. The facilitated excitation range is shaded by gray. Inset: The total steady population of multi-excitation onto states $|E^2R^1\rangle$ and $|E^3R^0
\rangle$ for $N=3$. Results in the presence(absence) of control atom are shown by the black(red-dashed) line. Rydberg population dynamics under the cases of (b1) no control atom and (b2) one control atom where $\Omega_p/\Omega_c = 0.4$. 
Similar behaviors with $\Omega_p/\Omega_c = 1.0$ and $2.0$ are displayed in (c1-c2) and (d1-d2). Populations $P_{g_c}(t)$, $P_{r_c}(t)$, $P_{g_c}(t)+P_{r_c}(t)$ are labeled by the black-solid line, the red-dashed line and the blue-solid line, respectively. } 
\label{fig2:steady}
\end{figure}

\section{Numerical Verification}

We next employ the above Monte Carlo method to study the realistic population dynamics in a coupled control-ensemble system. We consider the explicit case of $^{87}$Rb atoms trapped in two optical tweezers separated by $\bm{r}_0$, with energy levels $|g_{c(j)}\rangle=|5S_{1/2}\rangle$, $|e_{c(j)}\rangle=|5P_{3/2}\rangle$ and $|r_{c(j)}\rangle=|55S_{1/2}\rangle$. 
The corresponding interaction coefficient for state $|r_{c(j)}\rangle$ is $C_6/2\pi=50 $ GHz$\cdot\mu$m$^6$ and the non-fluctuated distance is $\bm{r_0}=3.062$ $\mu$m, where $\bm{r_0}>R_b=(\gamma_{eg}C_6/\Omega_c^2)^{1/6}\approx 1.6 $ $\mu$m is assumed for a {\it vdWs} interaction between the control-ensemble atom pairs \cite{PhysRevLett.110.203601}. At $T\approx 1$ $\mu$K the trap frequency is $\omega/2\pi=100$ kHz, leading to the standard deviation $\sigma = 14.3$ nm. The spontaneous decay rates are 
$\Gamma_e/2\pi=6.06$ MHz, $\Gamma_r/2\pi=2$ kHz, the dephasing rates are $\gamma_{re(eg)}/2\pi=12.12$ kHz and $\Omega_c/2\pi=6.06$ MHz, $\Delta/2\pi=121.2$ MHz.

In Fig. \ref{fig2:steady}(a), we show the steady-state Rydberg population $f_r$ of the targeted ensemble as a function of $\Omega_p/\Omega_c$(by varying $\Omega_p$) for $N=3$. When the control atom is unexcited we have $f_r = P_{g_c}(\infty)$; otherwise, we have $f_r = P_{g_c}(\infty)+P_{r_c}(\infty)$. For a weak probe field ($\Omega_p/\Omega_c < 0.5$), $f_r$ is almost same. That means the second two-photon process from $|g_cE^0R^1\rangle$ to $|r_cE^0R^1\rangle$ plays a negligible role and the system evolves samely in both cases. Yet 
once $\Omega_p/\Omega_c>0.5$, the steady Rydberg population can be clearly enhanced with the help of an excited control atom and quickly approaches 0.9204 at $\Omega_p/\Omega_c= 2$. While this value is only 0.7994 in the case of no control atom.
This enhancement effect comes from the second two-photon transition onto the double Rydberg state $|r_cE^0R^1\rangle$ enabled by the antiblockade condition which becomes more apparent in the case of a stronger probe field.
Other off-resonance collective states are almost unpopulated in our calculation. Especially, the population onto multi-excitation intermediate states keeps below $10^{-5}$, see the inset of Fig.\ref{fig2:steady}(a).

To further verify the above steady-state analysis, we study the time-dependent dynamics of the Rydberg-state populations in Fig.\ref{fig2:steady}(b1-d2), which visibly presents the population transfer among different collective states. From top to bottom we choose $\Omega_p/\Omega_c=(0.4,1.0,2.0)$. It is clear that with a weak probe driving $\Omega_p=0.4\Omega_c$, the excitation of the targeted ensemble mainly occupies state $|g_cE^0R^1\rangle$ with a small fraction of $P_{g_c}(t)$(black-solid) except the ground state(not shown), no matter whether the control atom is excited or not. By comparing (b1) and (b2) we find $P_{r_c}$(red-dashed) in (b2) is substantially suppressed agreeing with our theoretical prediction.
However for a stronger probe driving, the control atom's excitation would result in a subsequently new transition channel, which leads to an enhanced Rydberg probability redistributed on states $|g_cE^0R^1\rangle$ and $|r_cE^0R^1\rangle$. In (c2,d2) the population $P_{r_c}$(red-dashed) obtains a dramatic increases for a larger $\Omega_p$, arising the steady-state Rydberg population of ensemble atoms described by $P_{r_c}+P_{g_c}$(blue-solid) greatly enhanced. So we verify that the coupled control-ensemble system can exhibit a clear excitation facilitation effect for the targeted ensemble atoms when the probe field is not so weak.

\section{\label{sec:level1}Scheme Feasibility}
\subsection{$N$ ensemble atoms}

\begin{figure}
\centering\includegraphics[scale=0.4]{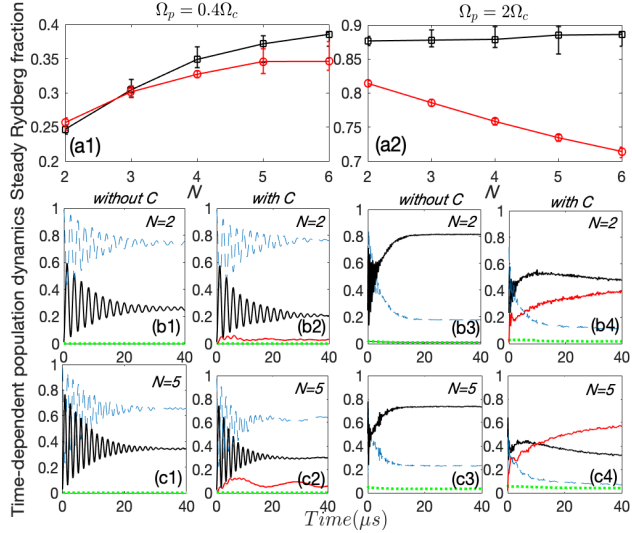}
\caption{(a1-a2) Steady-state Rydberg population $f_r$ {\it vs.} the change of atomic number $N$ in the ensemble. Results with(or without) the control atom are shown by the black line with squares(or the red line with circles). Every point is obtained by averaging over 300 stochastic realizations at $T=1$ $\mu$K and the error bar is given accordingly. (b1-b4) Time-dependent population dynamics for $N=2$ under the cases with (b1) $\Omega_p=0.4\Omega_c$ and no control atom; (b2) $\Omega_p=0.4\Omega_c$ and one control atom; (b3) $\Omega_p=2\Omega_c$ and no control atom; (b4) $\Omega_p=2\Omega_c$ and one control atom. $P_{g_c}(t)$, $P_{r_c}(t)$, $P_{g_cG}(t)$ are denoted by the black-solid line, the red-solid line and the blue-dashed line, respectively. Other intermediate-state population is labeled by the green-dotted line. (c1-c4) are same as (b1-b4) except for $N=5$.    }  
\label{fig3:atomnumber}
\end{figure}

We first discuss the excitation facilitation with different number of ensemble atoms. The experimental realization of a few-atom ensemble with a desired number of cold atoms could utilize a microscopic dipole trap \cite{PhysRevA.81.060308}. Depending on the local density of the MOT cloud around the trap, the exact number of trapped atoms can be desirably changed in experiment \cite{doi:10.1063/1.2206118}.

Figure \ref{fig3:atomnumber}(a1-a2) show the steady-state Rydberg population $f_r$ of the targeted atomic ensemble as a function of atomic number $N$. 
We choose $\Omega_p/\Omega_c=(0.4,2.0)$ representing the cases of weak and strong probe drivings.
As $N$ is increased, an intuitive feature is that $f_r$ slowly grows and approaches its saturation as shown in (a1) where a weak probe is applied. Because the probe Rabi frequency can obtain a usual $\sqrt{N}$-enhancement allowing for an enhanced steady-state Rydberg population \cite{PhysRevA.91.043802}. However, if beyond the weak probe limit [see (a2)] the steady Rydberg population $f_r$ exhibits an anomalous decrease. 
To understand this unusual effect
we revisit the collective states as shown in Fig.\ref{fig1:model}(a3) for single atomic ensemble. Under the strong probe limit state $|E^2R^0\rangle$ can be ignored due to its far-off-resonance detuning $2\Delta$. Therefore the Hamiltonian of the atomic ensemble can be rewritten by using four collective states
\begin{eqnarray}
{\mathcal{H}}_{en}^\prime&=&-\Delta(|E^1R^1\rangle\langle E^1R^1|+|E^1R^0\rangle\langle E^1R^0|)\nonumber\\
&+&(\sqrt{N}\Omega_p|G\rangle\langle E^1R^0|+\Omega_c|E^1R^0\rangle\langle E^0R^1|+\sqrt{N-1}\Omega_p|E^0R^1\rangle\langle E^1R^1|+\text{H.c.}).
\label{eq22}
\end{eqnarray}
When $\Delta \gg \Omega_p$ we safely ignore the doubly-excited state $|E^1R^1\rangle$ and ${\mathcal{H}}_{en}^\prime$ can be reduced into
\begin{equation}
\hat{\mathcal{H}}_{en}^{\prime\prime}=-\Delta|E^1R^0\rangle\langle E^1R^0|+\frac{(N-1)\Omega_p^2}{\Delta}|E^0R^1\rangle\langle E^0R^1|
+(\sqrt{N}\Omega_p|G\rangle\langle E^1R^0|+\Omega_c|E^1R^0\rangle\langle E^0R^1|+\text{H.c.}).
\label{eq23}
\end{equation}
Based on Eq.(\ref{eq23}), it is obvious that the energy shift related to state $|E^0R^1\rangle$ is proportional to $ N$ which means $|E^0R^1\rangle$ is highly-shifted when the atomic number $N$ is increased leading to a smaller Rydberg excitation probability. Numerical results(red line with circles in (a2)) also agree with the above discussions.
It is remarkable that, by placing a control atom nearby, $f_r$(black line with squares) can be kept at a high level with an arbitrary $N$ value see (a2). This effect is mainly contributed by the second two-photon process between $|g_cE^0R^1\rangle$ and $|r_cE^0R^1\rangle$, leading to an enhanced Rydberg population redistributed on these states. The large spontaneous loss from intermediate state $|g_cE^1R^0\rangle$ can be overcome by this facilitation effect.

Figure \ref{fig3:atomnumber}(b1-b4) and (c1-c4) display the population dynamics on three prominent states $|g_cG\rangle$, $|g_cE^0R^1\rangle$, $|r_cE^0R^1\rangle$ for $N=2$ and $5$. When the probe field is weak {\it i.e. }$\Omega_p=0.4\Omega_c$ in (b1,b2,c1,c2), a large fraction of population is observed to stay on the ground state $|g_cG\rangle$ (blue-dashed). But with the help of control atom, the Rydberg population $P_{r_c}$(red-solid) obtains a clear enhancement with the atomic number $N$. Therefore even in the weak probe regime, the excitation facilitation effect inside a control-ensemble system is easily observable as long as ensemble atoms are sufficient.
Most importantly, when the probe driving is strong $\Omega_p=2\Omega_c$, $f_r$ suffers from a strong decrease as $N$ increases. Hence, by comparing (b3) and (c3) we find $P_{g_c}$(black-solid) suffers from a reduction accompanied by the increase of $P_{g_cG}$(blue-dashed) on the ground state. However in the coupled control-ensemble system, benefiting from an enhanced Rydberg population which is redistributed on states $|g_cE^0R^1\rangle$ and $|r_cE^0R^1\rangle$, the leakage from intermediate states can be suppressed. The total Rydberg population $P_{g_c}+P_{r_c}$ of the atomic ensemble can be persistently kept at a high level with arbitrary $N$.
This means the controllable excitation facilitation requires a relatively strong probe strength and should be more observable in a realistic atomic ensemble with sufficient atoms.

\subsection{Thermal motion of atoms}

\begin{figure}
\centering\includegraphics[scale=0.4]{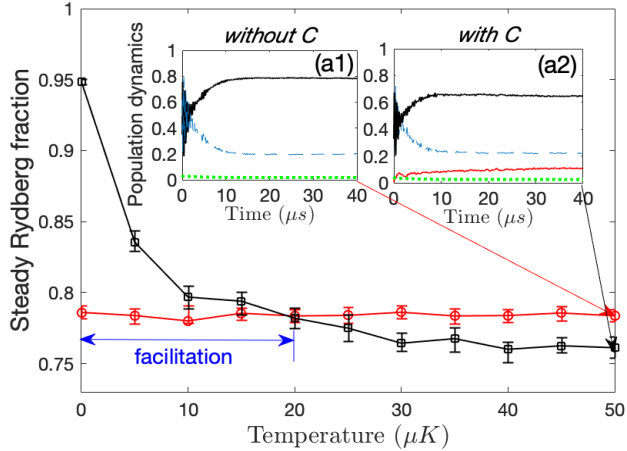}
\caption{Steady-state Rydberg population $f_r$ of the targeted ensemble {\it vs.} the atomic temperature $T$ for $N=3$. Results with(or without) the control atom are shown by the black line with squares(or the red line with circles). The excitation facilitation regime is shown in blue. Insets (a1-a2): The corresponding time-dependent population dynamics at $T=50$ $\mu$K. The linetype and parameters are same as used in Fig.\ref{fig2:steady} except for $\Omega_p=2\Omega_c$. }
\label{fig4:pertur}
\end{figure}

To provide a profound study for this effect we further investigate the behavior of population dynamics under different temperatures. Intuitively the increase of atomic temperature would easily break the antiblockade condition, leading to the fluctuated component $\delta U(\bm{r}_j)$ non-negligible \cite{Ostmann_2019}. As a consequence the doubly-excited Rydberg state $|r_cE^0R^1\rangle$ becomes off-resonance accompanied by a big reduction of the population $P_{r_c}$. 
A comparison for the behaviors of steady Rydberg populations {\it vs.} the temperature $T$, is illustrated in Fig.\ref{fig4:pertur}. 
By increasing $T$ it is apparent that $f_r$ almost keeps a constant in the case with no control atom due to the perfect blockade effect preserved inside the ensemble atoms. Because even at $T=50$ $\mu$K the distance deviation of ensemble atoms is only $\sigma \approx 0.101 $ $\mu$m, which is smaller than the blockade radius $R_b\approx 1.6$ $\mu$m by one order of magnitude.

As turning to the case with a control atom the steady-state Rydberg population reveals a clear decrease due to the influence of fluctuated interaction $\delta U(\bm{r}_j)$ that increases with $T$. \textcolor{black}{From Eq.(\ref{devr}) a rough estimation shows $\delta U(\bm{r}_j)/2\pi\approx -11.98$ MHz if $|\bm{r}_j-\bm{r}_0|=0.1$ $\mu$m at $T=50$ $\mu$K.} This value leads to a non-negligible off-resonance detuning to the doubly-excited Rydberg state $|r_cE^0R^1\rangle$, arising the breakdown of antiblockade facilitation $\delta\neq-U(\bm{r}_j)$. Therefore the controllable excitation facilitation merely exists at a relatively lower temperature. Based on our simulations when $T\leq20$ $\mu$K the effect should be observable within a few-atom ensemble containing three atoms \cite{PhysRevA.85.062708}. Since $f_r$ decreases with $N$ in a single-ensemble system [see Fig.\ref{fig3:atomnumber}(a2)], 
a wider regime of temperature for excitation facilitation can be obtained if more ensemble atoms are involved. So far a desired number of cold atoms can be prepared in an optical dipole trap \cite{PhysRevLett.85.3777} which makes this excitation facilitation effect more observable experimentally.
Realistic population dynamics as shown in Fig.\ref{fig4:pertur}(a1-a2) agrees with the above steady-state discussions, in which the population $P_{r_c}$(red-solid, (a2)) has been deeply suppressed due to the off-resonantly coupled state $|r_cE^0R^1\rangle$. Therefore no facilitation can be found at $T=50$ $\mu$K because $f_r$(=$P_{g_c}(\infty)+P_{r_c}(\infty)$ in (a2))$<f_r$(=$P_{g_c}(\infty)$ in (a1) ).

\subsection{Adjustable control-ensemble distances}

\begin{figure}
\centering\includegraphics[scale=0.42]{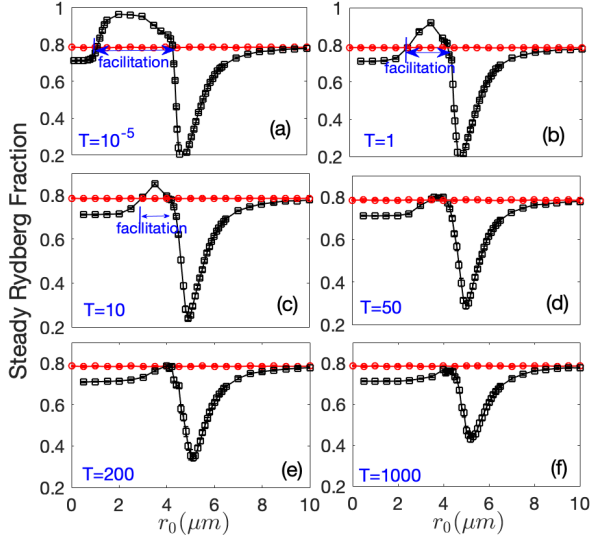}
\caption{(a-f) Steady-state Rydberg population $f_r$ of the targeted ensemble {\it vs.} the control-ensemble distance $\bm{r_0}$ under various $T$(in unit of $\mu$K), corresponding to the distance deviations $\sigma = (4.5\times10^{-5},0.014,0.045,0.101,0.202,0.451)$ $\mu$m. Linetypes and parameters are same as used in Fig.\ref{fig4:pertur}.} 
\label{fig5:distance}
\end{figure}

To manipulate the internal states of ensemble atoms, the assisted control atom which is individually addressed in a dipole trap, should keep an appropriate distance $\bm{r}_0$ from the ensemble trap. This relative distance is typically adjustable within a range of several $\mu$m via tuning the incidence angle of the focusing laser beams \cite{PhysRevLett.110.263201}. 
Here we study the influence on the steady-state Rydberg population of ensemble atoms by using a variable distance $\bm{r}_0$, equivalent to a variable $U(\bm{r}_0)$. It should be noted that the condition $\delta = -U(\bm{r}_0)$ set for calculation makes the detuning $\delta$ changeable with $\bm{r}_0$ at the same time.
Intuitively, the manipulation of control atom would vanish as $\bm{r}_0\to \infty$ and both the control and atomic ensemble behave independently then. And when $\bm{r}_0$ is adjusted within an appropriate range, the excitation of ensemble atoms must be affected.

 Figure \ref{fig5:distance}(a-f) show the steady-state Rydberg population as a function of the control-ensemble distance $\bm{r}_0$ under different atomic temperatures. It is evident that $f_r$ of single atomic ensemble is unvaried because of the preservation of perfect blockade effect, which agrees with the results in Fig.\ref{fig4:pertur}. While we note that in the vicinity of a control atom, by varying $\bm{r}_0$ there exists a clear dip $\bm{r}_0=\bm{r}_{dip}$ at which the $f_r$ attains its minimum. To explain this point, we revisit the full level scheme in Fig.\ref{fig1:model}(a). 
 At $\bm{r}_0=\bm{r}_{dip}$, another transition channel
 \begin{equation}
|g_cG\rangle\to|e_cG\rangle\to|r_cG\rangle\to|r_cE^1R^0\rangle\to|r_cE^0R^1\rangle
 \end{equation}
 is facilitated, where the effective detuning $\Delta_{r_cG}$ with respect to $|r_cG\rangle$ can be estimated as
\begin{equation}
    \Delta_{r_cG}=\frac{\Omega_c^2-\Omega_p^2}{\Delta}+\delta_{dip}+\frac{N\Omega_p^2}{\Delta-\delta_{dip}}.
\end{equation}
 By considering $\Delta\gg|\delta|$ and $\Delta_{r_cG}=0$, the steady Rydberg population eventually stays on state $|r_cG\rangle$ and the position $\bm{r}_{dip}$ can be analytically solved by
\begin{equation}
 \bm{r}_{dip}=(\frac{C_6}{-\delta_{dip}})^{1/6}\approx(\frac{C_6\Delta}{\Omega_c^2+(N-1)\Omega_p^2})^{1/6}.
 \label{deltadip2}
\end{equation}
 Notice that at $\bm{r}_0=\bm{r}_{dip}$, the doubly-excited Rydberg state $|r_cE^0R^1\rangle$ is off-resonance due to a non-zero effective detuning $\frac{\Omega_c^2}{\Delta-\delta_{dip}}-\frac{\Omega_p^2}{\Delta}\neq 0$. Hence the excitation of control atom prevents the ensemble atoms from excitation at this position. By using $\Omega_c=\Gamma_e=2\pi\times 6.06$ MHz, $\Delta=20\Gamma_e$, $\Omega_p/\Omega_c=2.0$ and $N=3$, it gives $\bm{r}_{dip}\approx 5.1$ $\mu$m, agreeing with the numerical results in Fig.\ref{fig5:distance}.

 In addition we find that, the excitation facilitation effect is observable only if $\bm{r}_0$ is appropriate. 
 Based on Fig.\ref{fig5:distance}, as expected there is no facilitation when $\bm{r}_0\to\infty$, where the ensemble atoms would obtain an independent excitation just like a single atomic ensemble. On the contrary if $\bm{r}_0\to 0$, a stronger fluctuated interaction $\delta U(\bm{r}_j)$ would result in a complete breaking of the 
 antiblockade condition by the control atom. Hence no excitation facilitation occurs when $\bm{r}_0$ is very small [see Fig.\ref{fig5:distance}(a-f)]. Only if the control-ensemble distance $\bm{r}_0$ is appropriately adjusted the excitation facilitation effect can take place.
  Moreover no facilitation exists when $T>50$ $\mu$K. Because at a higher temperature, the thermal motion of ensemble atoms strongly breaks the antiblockade condition, see more details in Sec.5.2.

 Finally it is worth stressing that an experimental observation of this excitation facilitation effect depends on an appropriate adjustment of both the control-ensemble distance and the atomic temperature, as well as the inclusion of sufficient ensemble atoms.

 \section{Conclusion}
 
 We propose a new scheme to realize excitation facilitation of a strongly-interacting Rydberg ensemble enabled by an auxiliary control atom. By utilizing a relatively strong probe driving, a simultaneously-excited control atom can counterintuitively add to a second two-photon transition onto the doubly-excited Rydberg state, which arises an excitation facilitation effect for
 the steady Rydberg population of ensemble atoms. Differing from the single-ensemble model where only the singly-excited Rydberg state can be occupied, this enhanced Rydberg probability would be redistributed on the singly- and doubly-excited Rydberg states under the antiblockade facilitation condition.
 By studying various experimentally-feasible parameters which set some constraints for this effect, we show the combination of a strong probe driving and a low temperature is an important prerequisite for observing excitation facilitation in a realistic Rydberg atomic ensemble.

 Our achievements, on one hand, arise a more profound understanding of the collective excitation feature of Rydberg-atom ensembles (see a review in \cite{Browaeys2020}); and on the other hand, are able to realize a flexible control for the facilitated excitation in a single atomic ensemble. The next step towards the use of this mechanism would be realizing a mesoscopic Rydberg controlled-phase gate \cite{Yin:21}. Because the excitation of control atom could add to a quasi-dark eigenstate between $|G\rangle$ and $|E^0R^1\rangle$ mediated by $|E^1R^0\rangle$. By varying the amplitudes of probe laser and the two-photon detuning, the system will experience an adiabatic evolution along this quasi-dark state which can accumulate an effective dynamic phase through the nonzero eigenenergy. A careful modulation of the laser profiles can be used for implementing a mesoscopic Rydberg gate with an arbitrary phase factor \cite{PhysRevApplied.18.044042,PhysRevApplied.13.024059,PhysRevA.101.062309}, which will leave for our future work.

\providecommand{\noopsort}[1]{}\providecommand{\singleletter}[1]{#1}%

\newpage

\end{CJK*}  

\begin{thebibliography}{10}
\newcommand{\enquote}[1]{``#1''}

\bibitem{PhysRevLett.114.203002}
~Urvoy A, ~Ripka F, ~Lesanovsky I, ~Booth D, Shaffer J ~P, ~Pfau T , and
~L\"ow R 2015 {{Phys. Rev. Lett.}} \textbf{114}, 203002
 

\bibitem{PhysRevA.96.041602}
~Guti\'errez R, ~Simonelli C, ~Archimi M, ~Castellucci F, ~Arimondo E,
  ~Ciampini D, ~Marcuzzi M, ~Lesanovsky I, and ~Morsch O, 2017 {{Phys. Rev. A}} \textbf{96}, 041602

\bibitem{PhysRevLett.123.203603}
~Lampen J, ~Duspayev A, ~Nguyen H, ~Tamura H, Berman P ~R , and ~Kuzmich A,
  2019 {{Phys. Rev. Lett.}} \textbf{123}, 203603

\bibitem{PhysRevResearch.2.043339}
~Stiesdal N, ~Busche H, ~Kumlin J, ~Kleinbeck K, B\"uchler H ~P, and
  ~Hofferberth S, 2020 {{Phys. Rev. Research}} \textbf{2}, 043339

\bibitem{PhysRevLett.99.260501}
~Brion E, ~M\o{}lmer K, and ~Saffman M 2007 {{Phys. Rev. Lett.}} \textbf{99}, 260501 

\bibitem{Zhang_2021}
Zhang Z Y, Ding D S, and Shi  B S 2021 {{Chinese Physics B}} \textbf{30}, 020307

\bibitem{Wu_2021}
~Wu X, ~Liang X, ~Tian Y, ~Yang F, ~Chen C, Liu Y C, Tey M ~K, and ~You L 
  2021 {{Chinese Physics B}} \textbf{30}, 020305

\bibitem{Chinese Phys.B.29.083202}
Li P C  and Chu S I 2020 {{Chinese Phys.B}} \textbf{29} 083202 

\bibitem{PhysRevA.98.043836}
~Khazali M 2018 {{Phys. Rev. A}} \textbf{98}, 043836 

\bibitem{Yang2022}
Yang C W, ~Yu Y, ~Li J, ~Jing B, Bao X H, and Pan J W,
  2022 {{Nature Photonics}} \textbf{16}, 658--661

\bibitem{PhysRevLett.128.060502}
Sun P F, ~Yu Y, An Z Y, ~Li J, Yang C W, Bao  X H, and Pan  J W 2022 {{Phys. Rev. Lett.}} \textbf{128}, 060502

\bibitem{Yang:22}
Yang C W, ~Li J, Zhou M T, ~Jiang X, Bao X H, and Pan J W
   2022 {{Optica}} \textbf{9}, 853--858 

\bibitem{PhysRevResearch.3.033287}
~Padr\'onBrito A, ~Lowinski J, ~Farrera P, ~Theophilo K, and ~de~Riedmatten H 2021 {{Phys. Rev. Research}} \textbf{3}, 033287

\bibitem{PhysRevA.103.023703}
~Petrosyan D and ~M\o{}lmer K 2021 {{Phys. Rev. A}} \textbf{103}, 023703 

\bibitem{PhysRevA.102.042607}
Guo C Y, Yan L L, ~Zhang S, Su S L, and ~Li W 2020 {{Phys. Rev. A}} \textbf{102}, 042607 

\bibitem{PhysRevA.98.062326}
Gujarati T ~P 2018 {{Phys. Rev. A}} \textbf{98}, 062326

\bibitem{PhysRevResearch.4.033087}
~Haase T, ~Alber G, and Stojanovi\ifmmode~\acute{c}\else \'{c}\fi{} V ~M 2022
  {{Phys. Rev. Research}} \textbf{4}, 033087 

\bibitem{RevModPhys.82.2313}
~Saffman M, Walker T ~G, and ~M\o{}lmer K 2010 {{Rev. Mod. Phys.}} \textbf{82},
  2313--2363

\bibitem{Motzoi_2018}
~Motzoi F and ~M{\o}lmer K 2018 {{New Journal of
  Physics}} \textbf{20}, 053029 

\bibitem{PhysRevLett.102.170502}
~M\"uller M, ~Lesanovsky I, ~Weimer H,  B\"uchler H ~P, and ~Zoller P 2009 {{Phys. Rev. Lett.}} \textbf{102}, 170502

\bibitem{PhysRevLett.99.163601}
~Heidemann R, ~Raitzsch U, ~Bendkowsky V, ~Butscher B, ~L\"ow R, ~Santos L, and
  ~Pfau T 2007 {{Phys. Rev. Lett.}}
  \textbf{99}, 163601

\bibitem{PhysRevA.87.053414}
~Petrosyan D, ~H\"oning M, and ~Fleischhauer M 2013 {{Phys. Rev. A}} \textbf{87}, 053414

\bibitem{PhysRevX.5.031015}
~Zeiher J, ~Schau\ss{} P, ~Hild S, ~Macr\`{\i} T, ~Bloch I, and ~Gross C
  2015 {{Phys. Rev. X}} \textbf{5}, 031015 

\bibitem{PhysRevLett.128.123601}
~Mei Y, ~Li Y, ~Nguyen H, Berman P ~R, and ~Kuzmich A 2022 {{Phys. Rev. Lett.}}
  \textbf{128}, 123601

\bibitem{PhysRevLett.107.213601}
~Petrosyan D, ~Otterbach J, and ~Fleischhauer M,  2011 {{Phys. Rev.
  Lett.}} \textbf{107}, 213601 

\bibitem{Dudin2012}
 Dudin Y ~O, ~Li L, ~Bariani F, and ~Kuzmich A 2012 {{Nature
  Physics}} \textbf{8}, 790--794 

\bibitem{PhysRevA.87.023401}
~H\"oning M, ~Muth D, ~Petrosyan D, and ~Fleischhauer M 2013
  {{Phys. Rev. A}} \textbf{87}, 023401 

\bibitem{PhysRevA.97.043811}
Tian X D, Liu Y M, Bao Q Q, Wu J H, ~Artoni M, and La~Rocca G ~C 2018 {{Phys. Rev. A}} \textbf{97}, 043811

\bibitem{PhysRevLett.105.193603}
Pritchard J ~D, ~Maxwell D, ~Gauguet A, Weatherill K ~J, Jones M ~P ~A, and
  Adams C ~S 2010 {{Phys. Rev. Lett.}} \textbf{105},
  193603

\bibitem{PhysRevApplied.19.014017}
~Ding Y, ~Bai Z, ~Huang G, and ~Li W 2023 {{Phys. Rev. Appl.}} \textbf{19}, 014017
  
\bibitem{Qiao_2021}
~Qiao C and ~Zhang W 2021 {{Journal
  of Physics B: Atomic, Molecular and Optical Physics}} \textbf{54}, 205501

\bibitem{PhysRevA.105.043715}
Berman P ~R, ~Nguyen H, and Rojo A ~G 2022 {{Phys. Rev. A}} \textbf{105}, 043715 

\bibitem{PhysRevA.89.033839}
Liu Y M , ~Yan D, Tian X D, Cui C L, and Wu J H 2014 {{Phys. Rev. A}} \textbf{89}, 033839

\bibitem{PhysRevLett.113.233002}
~G\"arttner M, ~Whitlock S, Sch\"onleber D ~W, and ~Evers J 2014 {{Phys. Rev. Lett.}} \textbf{113}, 233002

\bibitem{Bai_2018}
Bai S Y, Bao Q Q, Tian X D, Liu Y M , and Wu J H 2018
  {{Journal of Physics B: Atomic, Molecular and Optical
  Physics}} \textbf{51}, 075502

\bibitem{PhysRevLett.98.023002}
~Ates C, ~Pohl T, ~Pattard T, and  Rost J ~M 2007 {{Phys. Rev. Lett.}} \textbf{98}, 023002

\bibitem{PhysRevLett.104.013001}
~Amthor T, ~Giese C, Hofmann C ~S, and ~Weidem\"uller M 2010 {{Phys. Rev.
  Lett.}} \textbf{104}, 013001

\bibitem{Kara2018}
~Kara D, ~Bhowmick A, and Mohapatra A ~K 2018 {{Scientific Reports}} \textbf{8}, 5256 

\bibitem{Bai_2020}
~Bai S, ~Tian X, ~Han X, ~Jiao Y, ~Wu J, ~Zhao J, and ~Jia S 2020 {{New Journal of Physics}}
  \textbf{22}, 013004 

\bibitem{Petrosyan_2013}
~Petrosyan D 2013 {{Journal of Physics B:
  Atomic, Molecular and Optical Physics}} \textbf{46}, 141001 

\bibitem{PhysRevLett.87.037901}
Lukin M ~D,  ~Fleischhauer M, ~Cote R, Duan L ~M, ~Jaksch D, Cirac J ~I, and
  ~Zoller P 2001 {{Phys. Rev. Lett.}}
  \textbf{87}, 037901

\bibitem{doi:10.1126/science.1217901}
Dudin Y ~O and ~Kuzmich A 2012 {{Science}} \textbf{336},
  887--889 

\bibitem{Carmele_2014}
~Carmele A, ~Vogell B, ~Stannigel K, and ~Zoller P 2014 {{New Journal of Physics}} \textbf{16}, 063042

\bibitem{PhysRevA.87.023827}
~Yan D, Cui C L, Liu Y M, Song  L J, and Wu J H 2013 {{Phys. Rev. A}}
  \textbf{87}, 023827

\bibitem{PhysRevA.86.023828}
~Yan D, Liu Y M, Bao Q Q, Fu C B, and Wu J H 2012 {{Phys. Rev. A}}
  \textbf{86}, 023828

\bibitem{doi:10.1080/23746149.2022.2064231}
Andersen M ~F 2022 {{Advances in Physics: X}} \textbf{7}, 2064231

\bibitem{PhysRevLett.96.063001}
Yavuz D ~D, Kulatunga P ~B, ~Urban E, Johnson T ~A, ~Proite N, ~Henage T,
 Walker T ~G, and ~Saffman M 2006 {{Phys.
  Rev. Lett.}} \textbf{96}, 063001

\bibitem{PhysRevLett.127.063604}
Spong N ~L ~R , ~Jiao Y, Hughes O ~D ~W ,Weatherill  K ~J, ~Lesanovsky I, and
 Adams  C ~S 2021 {{Phys. Rev. Lett.}} \textbf{127}, 063604

\bibitem{Ding_2019}
Ding Z X, Hu C S, Shen L T, Yang Z B, ~Wu H, and Zheng S B 2019 {{Laser Physics Letters}} \textbf{16}, 045203

\bibitem{PhysRevA.82.062328}
~Zuo Z and ~Nakagawa K 2010 {{Phys. Rev. A}} \textbf{82}, 062328

\bibitem{PhysRevA.101.042328}
~Li R, ~Yu D, Su S L, and ~Qian J 2020 {{Phys. Rev. A}} \textbf{101}, 042328

\bibitem{PhysRevLett.118.063606}
~Marcuzzi M, Min\'a\ifmmode~\check{r}\else \v{r}\fi{} J ~c ~v, ~Barredo D,
  x~de~L\'es\'eleuc D, ~Labuhn H, ~Lahaye T, ~Browaeys A, ~Levi E, and
  ~Lesanovsky I 2017 {{Phys. Rev. Lett.}} \textbf{118}, 063606

\bibitem{PhysRevLett.128.013603}
~Liu F, Yang Z C, ~Bienias P, ~Iadecola T, and Gorshkov A ~V 2022 {{Phys. Rev. Lett.}} \textbf{128}, 013603 

\bibitem{PhysRevA.98.062714}
~de~Hond J, ~van Bijnen R, Kokkelmans S ~J J M ~F, Spreeuw R ~J ~C, van~den Heuvell H ~B 
  v ~L, and van Druten N ~J 2018 {{Phys. Rev. A}} \textbf{98}, 062714

\bibitem{PhysRevA.91.043402}
~Petrosyan D, ~Rao D, and ~M\o{}lmer K 2015 {{Phys. Rev. A}} \textbf{91}, 043402

\bibitem{PhysRevA.85.063822}
Lee T ~E and Cros M ~C 2012 {{Phys. Rev. A}} \textbf{85},
  063822 

\bibitem{PhysRevLett.108.023602}
Lee T ~E, ~H\"affner H, and Cross M ~C 2012 {{Phys. Rev. Lett.}} \textbf{108},
  023602

\bibitem{PhysRevLett.110.203601}
Hofmann C ~S, ~G\"unter G, ~Schempp H, ~Robert-de Saint-Vincent M,
  ~G\"arttner M, ~Evers J, ~Whitlock S, and ~Weidem\"uller M 2013 {{Phys. Rev. Lett.}} \textbf{110}, 203601

\bibitem{PhysRevA.81.060308}
~Kruse J, ~Gierl C, ~Schlosser M, and ~Birkl G 2010 {{Phys. Rev. A}} \textbf{81},
  060308

\bibitem{doi:10.1063/1.2206118}
~Yoon S, ~Choi Y, ~Park S, ~Kim J, Lee J H, and ~An K 2006 {{Applied Physics Letters}}
  \textbf{88}, 211104.

\bibitem{PhysRevA.91.043802}
Liu Y M, Tian X D, ~Yan D, ~Zhang Y, Cui C L, and Wu J H 2015 {{Phys. Rev. A}}
  \textbf{91}, 043802

\bibitem{Ostmann_2019}
~Ostmann M, ~Marcuzzi M, ~Min{\'{a}}r J, and ~Lesanovsky I 2019 {{Quantum Science and Technology}} \textbf{4}, 02LT01

\bibitem{PhysRevA.85.062708}
~Fuhrmanek A, ~Bourgain R, Sortais Y ~R ~P, and ~Browaeys A 2012 {{Phys. Rev. A}} \textbf{85}, 062708

\bibitem{PhysRevLett.85.3777}
~Frese D, ~Ueberholz B, ~Kuhr S, ~Alt W, ~Schrader D, ~Gomer V, and
  ~Meschede D 2000 {{Phys. Rev.
  Lett.}} \textbf{85}, 3777--3780

\bibitem{PhysRevLett.110.263201}
~B\'eguin L, ~Vernier A, ~Chicireanu R, ~Lahaye T, and ~Browaeys A 2013 {{Phys. Rev. Lett.}} \textbf{110}, 263201 

\bibitem{Browaeys2020}
~Browaeys A and ~Lahaye T 2020 {{Nature Physics}}
  \textbf{16}, 132--142 

\bibitem{Yin:21}
Yin H D and Shao X Q 2021 {{Opt. Lett.}} \textbf{46}, 2541--2544 

\bibitem{Xu2022}
~Xu J and ~Liu T 2022 {{Quantum Information Processing}} \textbf{21}, 201
  
\bibitem{PhysRevLett.102.240502}
~Saffman M and ~M\o{}lmer K 2009 {{Phys. Rev. Lett.}}
  \textbf{102}, 240502

\bibitem{PhysRevApplied.18.044042}
~Li X, ~Shao X, and ~Li W 2022 {{Phys. Rev. Appl.}} \textbf{18}, 044042

\bibitem{PhysRevApplied.13.024059}
~Sun Y, ~Xu P, Chen P X, and ~Liu L 2020 {{Phys. Rev. Appl.}} \textbf{13}, 024059

\bibitem{PhysRevA.101.062309}
~Saffman M, Beterov I ~I, ~Dalal A, P\'aez E ~J, and Sanders B ~C 2020 {{Phys. Rev. A}} \textbf{101}, 062309

\end{thebibliography}
\end{document}